\documentclass[sigconf, screen]{acmart}
\AtBeginDocument{%
  \providecommand\BibTeX{{%
    \normalfont B\kern-0.5em{\scshape i\kern-0.25em b}\kern-0.8em\TeX}}}
    
\setcopyright{acmcopyright}
\copyrightyear{2023} 
\acmYear{2023} 
\setcopyright{rightsretained} 
\acmConference[EASE '23]{Proceedings of the International Conference on Evaluation and Assessment in Software Engineering}{June 14--16, 2023}{Oulu, Finland}
\acmBooktitle{Proceedings of the International Conference on Evaluation and Assessment in Software Engineering (EASE '23), June 14--16, 2023, Oulu, Finland}\acmDOI{10.1145/3593434.3593460}
\acmISBN{979-8-4007-0044-6/23/06}
  
\usepackage{acronym} 
\usepackage{enumitem}
\usepackage{amsmath}

\definecolor{ultramarine}{rgb}{0.7882, 0.2863,  0.0196}
\definecolor{specialblue}{rgb}{0.05, 0.40, 0.55}
\definecolor{sohistBlue}{rgb}{0.04, 0.36, 0.63}
\newcommand{\circled}[1]{\textcolor{sohistBlue}{\ensuremath{\text{\textcircled{\small #1}}}}}

\begin{document}
\acrodef{asat}[ASAT]{Automatic Static Analysis Tools}
\acrodef{td}[TD]{Technical Debt}
\acrodef{cd}[CD]{Code Debt}
\title[SoHist: A Tool for Managing Technical Deb]{\textit{SoHist:} A Tool for Managing Technical Debt through Retro Perspective Code Analysis}

\author{Benedikt Dornauer}
\additionalaffiliation{
    \institution{University of Cologne}
    \city{Cologne}
    \postcode{51147}
    \country{Germany} 
     }
\orcid{0000-0002-7713-4686}
\email{benedikt.dornauer@uibk.ac.at}
     
\author{Michael Felderer}
\additionalaffiliation{
    \institution{German Aerospace Center (DLR)}
    \department{Institute for Software Technology}
    \city{Cologne}
    \postcode{50923}
    \country{Germany}  
     }
\additionalaffiliation{
    \institution{University of Cologne}
    \city{Cologne}
    \postcode{51147}
    \country{Germany} 
     }    
\email{michael.felderer@uibk.ac.at}
\orcid{0000-0003-3818-4442}
\affiliation{%
  \institution{University of Innsbruck}
  \city{Innsbruck}
  \country{Austria}
  \postcode{6020}
}

\author{Johannes Weinzerl}
\orcid{0009-0009-2016-8584}
\email{jweinzerl@cccom.at}
\author{Mircea-Cristian Racasan}
\orcid{0009-0008-7938-3126}
\email{mracasan@cccom.at}
\affiliation{%
  \institution{c.c.com Moser GmbH}
  \streetaddress{Teslastraße 4, Grambach}
  \city{Graz}
  \country{Austria}}

\author{Martin Hess}
\orcid{0000-0001-5827-2736}
\email{martin.hess@softwareag.com}
\affiliation{%
  \institution{Software AG}
  \streetaddress{Uhlandstrasse 12}
  \city{Darmstadt}
  \postcode{64297}
  \country{Germany}
}

\renewcommand{\shortauthors}{Dornauer, et al.}
\begin{abstract}
    Technical debt is often the result of Short Run decisions made during code development, which can lead to long-term maintenance costs and risks. Hence, evaluating the progression of a project and understanding related code quality aspects is essential. 
    
    Fortunately, the prioritization process for addressing technical debt can be expedited with code analysis tools like the established SonarQube. Unfortunately, we experienced some limitations with this tool and have had some requirements from the industry that were not yet addressed. 
    
    Through this experience report and the analysis of scientific papers, this work contributes: (1) a reassessment of technical debt within the industry, (2) considers the benefits of employing SonarQube as well as its limitations when evaluating and prioritizing technical debt, (3) introduces a novel tool named \textit{SoHist} which addresses these limitations and offers additional features for the assessment and prioritization of technical debt, and (4) exemplifies the usage of this tool in two industrial settings in the ITEA3 SmartDelta project.      
\end{abstract}



\received{9 March 2023}
\received[revised]{22 April 2023}

\maketitle
\section{Introduction}
    \label{sec:introduction}
\begin{figure}[ht]
          \centering
          \includegraphics[width=\linewidth]{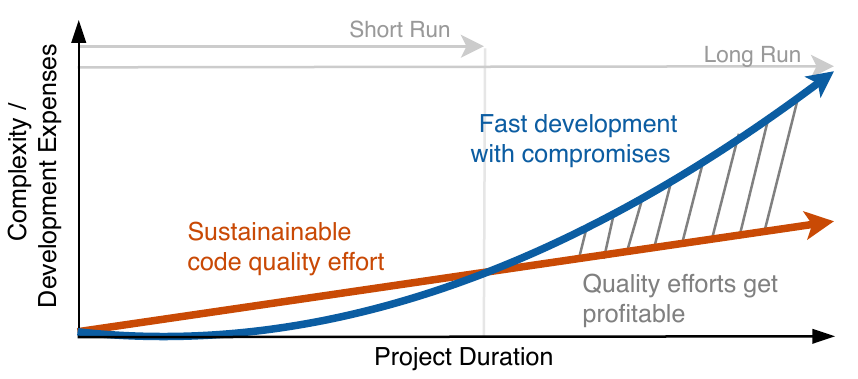}
          \caption{In a \textit{Short Run}, more features of software systems can emerge and lead to a better turnover. But if the code quality approaches are not “sustainable”, it may result in extra effort and overhead in the \textit{Long Run}, which could lead to total project failure.}     
          \label{fig:technicalProjectEvolvement}
    \end{figure}

    Nowadays, software development projects tend to focus more on the quick delivery of new features to compete in the highly competitive software market rather than on delivering high-quality code. While this approach may yield increased revenue in the \textit{Short Run}, it often results in high maintenance costs in the \textit{Long Run}. These costs are known as \ac{td} and are a severe problem in software development projects, as exemplified in Figure \ref{fig:technicalProjectEvolvement}. Numerous studies have substantiated these trends in several years of research~\cite{baskervilleHowInternetSoftware2001, vernerWhatFactorsLead2008b, ramacPrevalenceCommonCauses2022}, highlighting the detrimental impact of such circumstances on quality assurance.
    
    To counteract this trend, various factors have to be taken into account when evaluating, measuring, and finally addressing \ac{td}, which depend on the individual requirements and priorities of the project. Some projects may prioritize security, while others focus on performance scalability or test coverage to catch potential bugs and errors before they are deployed to production. Unfortunately, research on prioritizing \ac{td} is still in an early stage and lacks consensus on identifying important factors of \ac{td} and determining appropriate measurement methods \cite{baldassarreDiffusenessTechnicalDebt2020}.
    
    In the ITEA3 project \textit{SmartDelta}\footnote{\url{www.smartdelta.org} with partners \url{www.softwareag.com} and \url{www.cccom.at}}, e.g., we observed that some use case providers require a deeper understanding of how their projects have evolved over time in order to analyze and understand \ac{td}.  Among those are \textit{c.c.com Moser GmbH} and \textit{Software AG}. For this reason, we were looking for a tool that is capable to analyze code quality but also supports the historical analysis of code commits and deltas to investigate the code quality trend over time and to identify those code parts requiring the most maintenance effort.
    
\section{SonarQube: Advantages and Limitations for Analyzing Code Evolution}
    \label{sec:sonarqube}
\begin{figure}[ht]
          \centering
          \includegraphics[width=\linewidth]{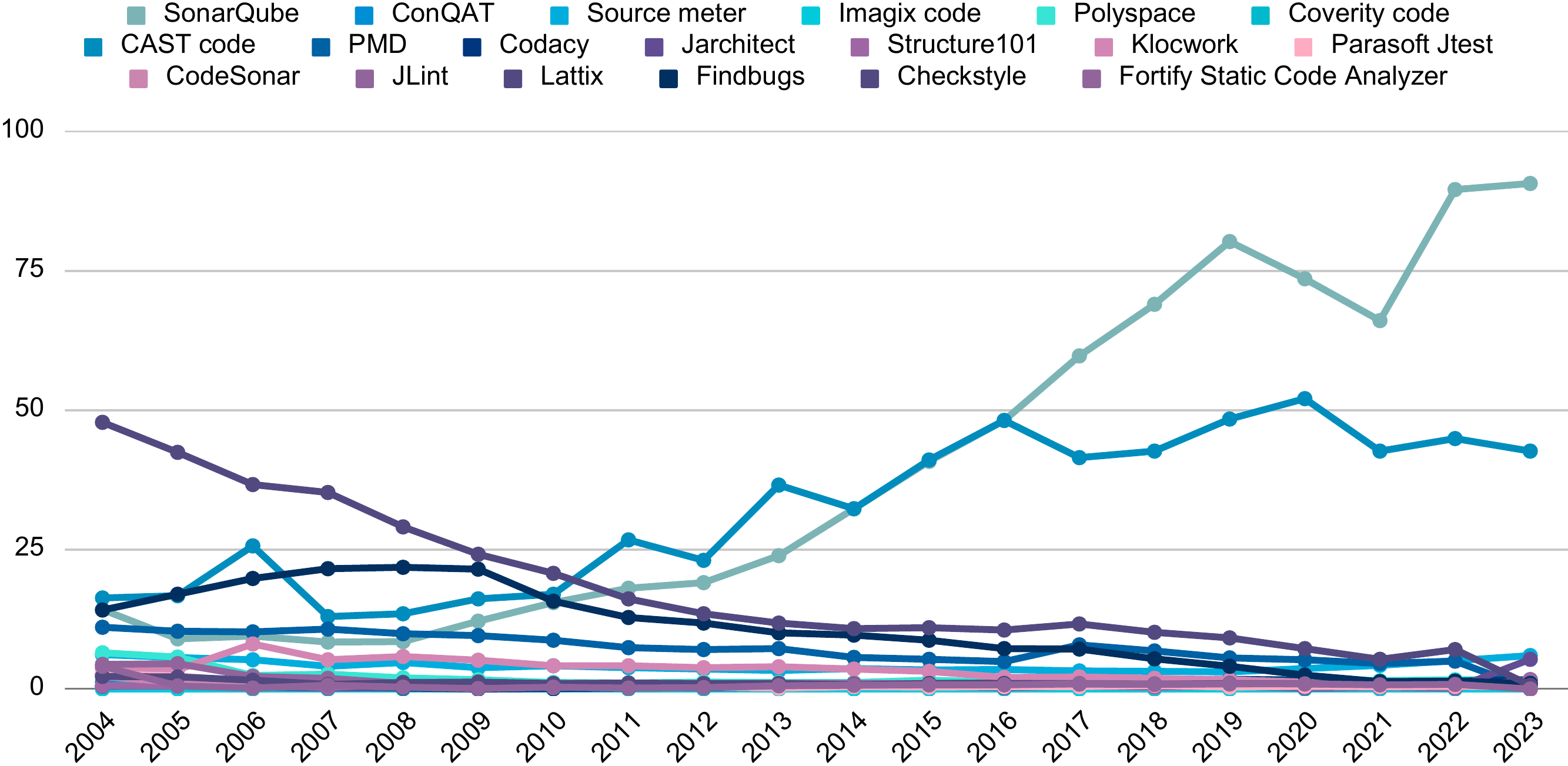}
          \caption{Google Search interest level in percent measured for popular code analysis tools since 2004. SonarQube shows the highest Google Search interest level since 2016 \cite{dornauerTrendCodeAnalysis2023}.} 
          \label{fig:trendCodeAnalysisTools}
    \end{figure}

    Various code quality analysis tools have been developed to support the improvement of reliability and quality in software systems. In 2018 Lenarduzzi et al.~\cite{lenarduzziSurveyCodeAnalysis2020} compiled a list of the most popular code analysis tools. Based on their findings, we adopted their methodology and investigated the current popularity of those tools in 2023. Our analysis ~\cite{dornauerTrendCodeAnalysis2023} shows that SonarQube\footnote{\url{www.sonarsource.com/products/sonarqube}} is the most popular among those (see Figure \ref{fig:trendCodeAnalysisTools}). 
    
    With more than 200 000 companies, as claimed by Sonar Source \cite{sonarsourceSonarQubeDownloads2023a}, the \textit{SonarQube Community Edition} is widely established in the industry \textcolor{specialblue}{[Adv.1]}. SonarQube gives companies, such as Software AG and c.c.com Moser GmbH, the opportunity to inspect code quality and code security in their software engineering pipeline, supporting over 19 programming languages in the free version and further 5 in the developer edition \textcolor{specialblue}{[Adv.2]}\cite{sonarsourceSonarQube}.
    
    The classification system for issues used by SonarQube can be grouped into three main types. The first category is \textit{Code Smells}, which are issues that can increase change-proneness and maintenance efforts. The second category is \textit{Bugs}, which are issues that could result in an error or unexpected behaviour. The third category is \textit{Vulnerabilities}, which identifies problems that can compromise the security of the software \cite{sonarsourceMetricDefinition}. All these issues compiled into one tool are a convenient and time-saving solution for industry \textcolor{specialblue}{[Adv.3]}.
    
    However, SonarQube also has some limitations, which we will discuss further: \\ 
    \textbf{\textcolor{ultramarine}{[Lim.1]}} \textbf{Focus on the latest project version}\\
    SonarQube analyses the modifications made to a project from the point of its initial integration. Consequently, the previous project's evolution might not be accessible, preventing the user from analysing the project's quality trend over time \cite{spencerHaveSonarQubeCreate2022}. Specifically, this could be of interest if someone is tasked with taking over a new project. Notably, there is a workaround, but it is complicated to implement, requires detailed knowledge, and is far from being fully automated.
    \noindent\textbf{\textcolor{ultramarine}{[Lim.2]}} \textbf{Limited analysis options} \\
    Another limitation of SonarQube Community Edition is the need for additional filtering options for more detailed assessments. For instance, it is not possible to analyze a different branch than the main branch or to focus the analysis on specific developers.\\  
    \textbf{\textcolor{ultramarine}{[Lim.3]}} \textbf{Lacking comparability between SonarQube versions}
    \begingroup
    Quality metrics and rules may change between different versions of SonarQube \cite{buchwaldHowOftenAre2022}. Consequently, such discrepancies can result in different outcomes and a distorted picture of the \ac{td} evolution depending on the SonarQube versions used during the project's lifetime. \\
    \textbf{\textcolor{ultramarine}{[Lim.4]}} \textbf{Reduced visualization capabilities}\\
    SonarQube offers some basic visualizations supporting three metrics to be shown at the same time. This requires, however, that the values of the selected metrics are in the same range because otherwise, smaller distinctions are no longer visible. Another aspect to consider is the need for a single metric for interpreting \ac{td}, which may vary depending on a case-by-case basis (see Section \ref{sec:introduction}). Unfortunately, SonarQube does not currently offer such a visualization.
    
\section{SoHist}
    Considering the industrial relevance of SonarQube in 2023 and the demand for extended (historical) analysis functionality, we decided to develop a new code analytics tool called \textit{SoHist} that addresses the limitations of SonarQube and provides extended historical code analysis capabilities such as the evolution of \ac{td} over time. SoHist is open-source and available on GitHub \texttt{bdornauer/sohist}\footnote{\url{https://github.com/bdornauer/sohist}}. 

\subsection{The Concept of SoHist}
    SoHist is a toolset available in a containerized format and can be deployed using \textit{Docker}. The system's architectural design encompasses a SonarQube instance along with a database, the SonarScanner to trigger the code analysis, an interface to Git, as well as a web service hosting SoHist's user interface. Figure \ref{fig:sohist_concept} illustrates the automated procedure for conducting a historical analysis of a Git repository history \textcolor{ultramarine}{[Lim.1]}.
        
    \begin{figure}[ht]
      \centering
      \includegraphics[width=0.83\linewidth]{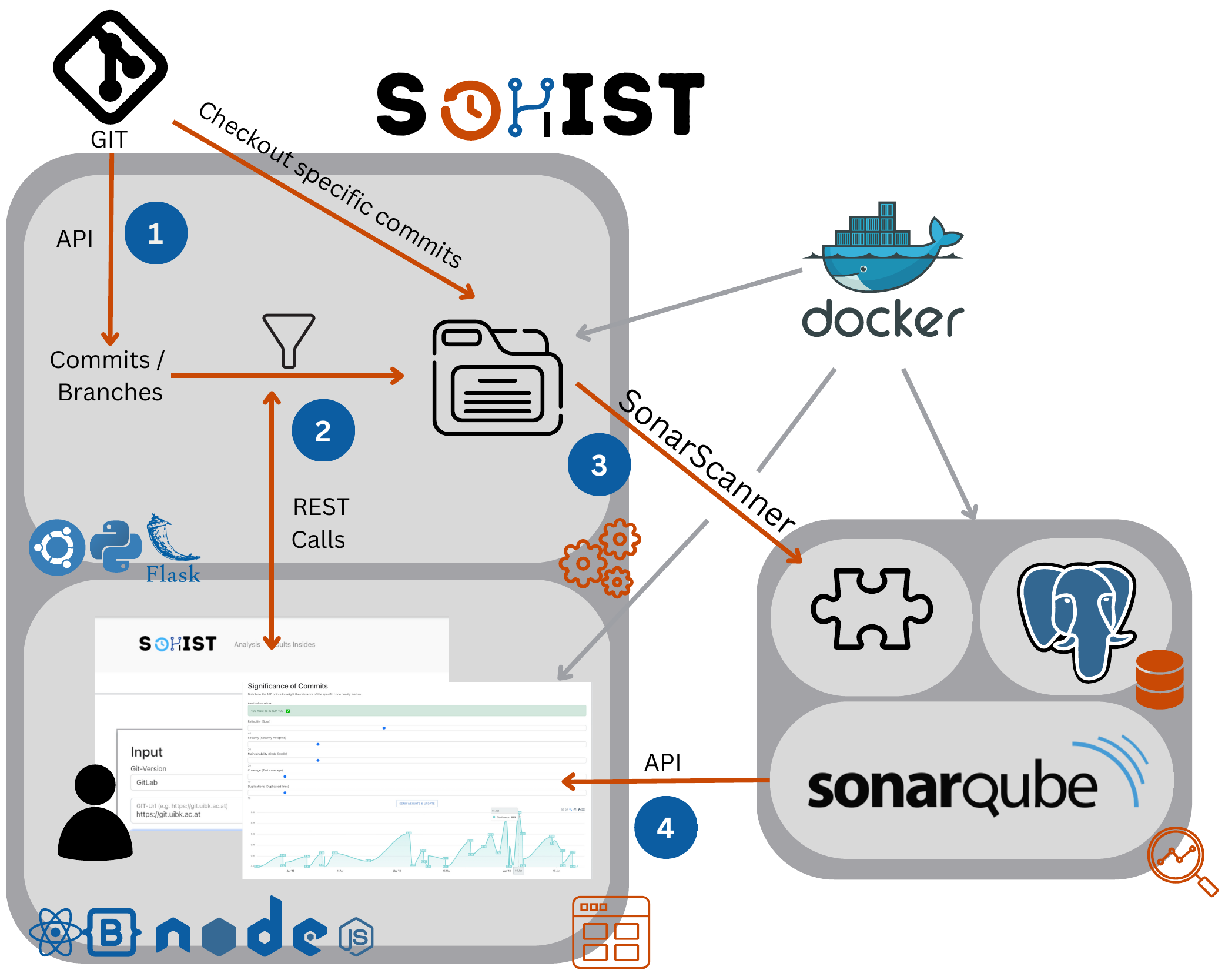}
      \caption{Conceptual structure of SoHist.}  
      \label{fig:sohist_concept}
    \end{figure}
    
    \circled{1} In the first step, the user must enter some required parameters to connect SoHist to the version control system. This includes, e.g., the URL of the target GitLab instance, the target project name, and the user's access token. 
    \circled{2} After connecting SoHist to the specified GitLab, the user can define the parameters for the historical code analysis of the selected project. This includes the time range of the analysis, the desired committers (and their corresponding commits), and a Git branch of their choice \textcolor{ultramarine}{[Lim.2]}. 
    \circled{3} Subsequently, SoHist analysis can be triggered by the user, which automatically executes individual SonarQube analysis runs for each change. By using the same SonarQube versions and thus identical metrics and rules for each run, SoHist ensures that the results for the changes are comparable \textcolor{ultramarine}{[Lim.3]}. If a newer SonarQube version is available, the retro perspective analysis with SoHist can be executed again, which could be cumbersome, but ensures comparability.   
     \circled{4} After the execution of SonarQube analysis, the user can access the code quality history and use the two available SoHist visualizations\footnote{Videos for demonstration: \url{https://doi.org/10.5281/zenodo.7713782}} -- \textit{Code Evolution} and \textit{Weighted Code Evolution Significance}. 
     This enables the user to track the evolution of the code quality over time and evaluate the impact of individual changes. 
    
\subsection{Outcomes and Visualization \textcolor{ultramarine}{[Lim. 4]}}
     SoHist provides the user with the capability to view multiple metrics simultaneously (Visualization 1: \textit{Code Evolution}). When the user moves the cursor over a specific time, the corresponding timestamp is highlighted across all metric charts. This allows the user to compare the various metrics against each other.
    \begin{figure}[ht]
          \centering
          \includegraphics[width=\linewidth]{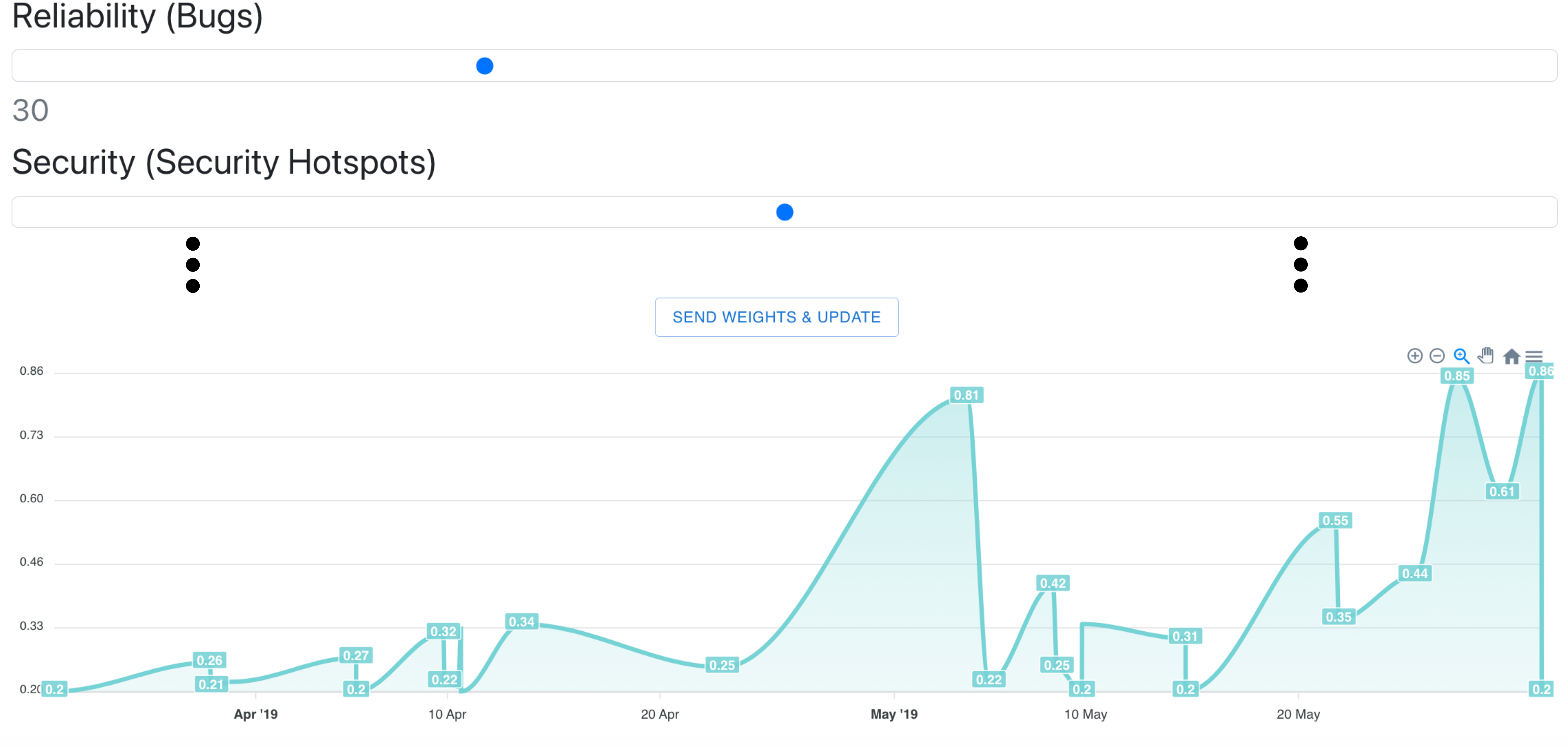}
          \caption{The user can prioritize specific metrics of interest in assigning weights. In this particular scenario of a university teaching project, Security issues were assumed to be critical (50\%), followed by Bugs (30\%) and Code Smells (20\%). Through this visualization, it can be observed that during the initial phase, when only experienced researchers were involved in setting up the repository (skeleton), there were fewer issues. Later on, as students joined, the number of issues focusing on security increased comparatively.} 
          \label{fig:weight_view}
    \end{figure}\\
    The second visualization called \textit{Weighted Code Evolution Significance} introduces a novel approach to address the challenge of individual project demands and prioritization of specific SonarQube's main metrics, as outlined in Section \ref{sec:sonarqube}. 
    This approach allows users to focus their analysis on multiple specific main metrics of their choice (Reliability, Security, Maintainability, Test Coverage and Duplicated Lines) by assigning individual metric weights.
    Based on this weighting and the data obtained from the individual SonarQube runs, a visual representation of the project's life cycle is shown, as demonstrated in Figure \ref{fig:weight_view}, highlighting those changes that have had the greatest impact according to the chosen weights.
    The higher the \textit{Weighted Code Evolution Significance}, the more significant the change related to the weighted category is. For a detailed description of the calculations, take a look at \cite{dornauerComputationsWeightedCode2023}. Further, a detailed analysis can be conducted using the first visualization. 
    
\section{Industrial Challenges and Usage of SoHist}
\label{sec:useCases}
  \subsection{c.c.com Moser GmbH - Logistics and Personal Mobility }
    \noindent\textbf{Background:} One of the solutions provided by c.c.com Moser GmbH are their BLIDS sensors, which enable the measurement of traffic flow on the street using Bluetooth, WiFi and Bluetooth Low Energy data. In temporary deployment scenarios, these sensors typically rely on battery power and require periodic replacement. Consequently, c.c.com Moser GmbH aims to reduce the energy drain of the sensors.
    
    \noindent\textbf{Scope:} Given the high cost associated with hardware exchange, c.c.com Moser GmbH  has undertaken efforts to address energy consumption at the software level (design decisions, data compression, etc.). The company has sought to establish potential correlations between software quality metrics and physical properties, such as specific energy consumption. Over the years, c.c.com Moser GmbH  has amassed a wealth of physical measurements that will be used in the upcoming analysis tasks. On the software level, c.c.com Moser GmbH  still needs to collect software quality metrics over time and has needed a tool that facilitates historical code analysis with support for multiple programming languages.
    
    \noindent\textbf{Usage of SoHist:} Therefore, c.c.com has initiated an analysis of the firmware repository. For every historical update of the sensor software, c.c.com collects energy-related code metrics via the SonarQube API. This is only feasible with SoHist's retrospective analysis, as no previous measurements are available. Metrics of particular interest include, for instance, the \textit{McCabe Cyclomatic Complexity} or \textit{Lines of Code} as prior research by Corrêa et al.~\cite{5578300} has demonstrated their potential impact on the energy consumption of embedded systems. In the ongoing activities of SmartDelta, c.c.com has the historical energy-related code metrics available and can investigate if specific changes (related to the code metrics) impacted the energy consumption. With SoHist, c.c.com Moser GmbH can now address its correlation efforts. 
    
    \subsection{Software AG – Enterprise Software}
    \textbf{Background:} Enterprises are highly dependent on the availability and reliability of their software in today’s world. Any malfunction resulting in unavailable or severely slowed down software systems may severely impact their business. Continuously improving the codebase is thus critical for Software AG as a leading supplier of enterprise software to deliver high-quality products and increase customer satisfaction.
    
    \noindent\textbf{Scope:} The more complex a software product gets, the more time and effort must be invested in code maintenance, leaving less time for feature development. Thus, identifying which parts of the code are causing problems – such as security or performance issues – or require increased maintenance is critical to efficiently solving issues and managing efforts and costs. 
    Software AG Research uses SonarQube for the detection of common bugs, security issues and for analyzing the overall code quality. However, SonarQube does not give direct insights into the historical evolution of the codebase. This information would enable the analysis of \ac{td} and could lead to a better understanding of a software development project's progression over time.
    
    \noindent\textbf{Usage of SoHist:} Software AG Research used SoHist to assess the code quality trend within one of their research projects. The SoHist analysis and especially the novel \textit{Weighted Code Evolution Significance} view, focusing on Bugs and Code Smells, revealed some interesting insights. Figure \ref{fig:sag_eval} presents the analysis results of the Master branch. The increased number of Bugs and Code Smells shown for the Master branch in the project's early to mid-stage \textcolor{sohistBlue}{(1)} are indicators for technical debt. Further analysis revealed that during this time frame, the team focused on developing new features rather than improving the code quality on the Master branch. In the middle of March, the team incorporated a new process for addressing code quality issues. This led to a substantial reduction in code quality issues. Thanks to the new process, the code quality issues observed at the end of May \textcolor{sohistBlue}{(2)} have been addressed and solved faster this time. Through SoHist, Software AG Research was able to reproduce this time frame and assess the actual benefits of the new quality assurance process. 
    \begin{figure}[ht]
          \centering
          \includegraphics[width=\linewidth]{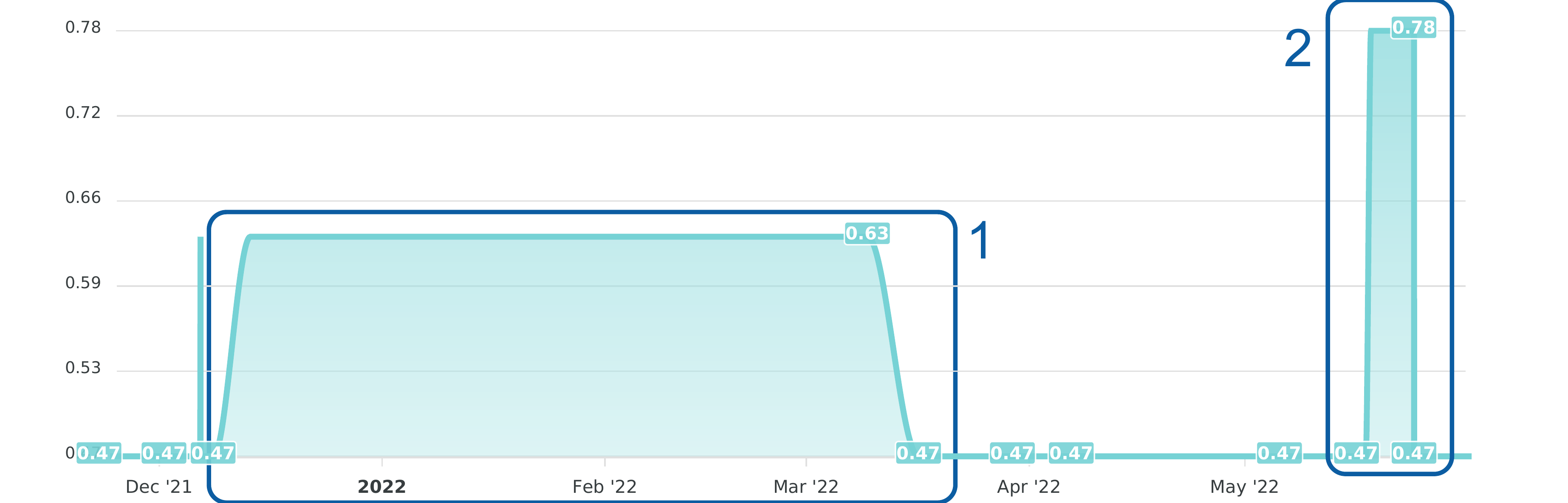}
          \caption{Software AG Use Case: Historical analysis of the master branch of a project, focusing on bugs and code smells.}  
          \label{fig:sag_eval}
    \end{figure}
\section{Conclusion}
    Maintaining a balance between developing new features and ensuring software quality is essential for the success of software development projects. Achieving this balance requires precise monitoring and management of \ac{td} overtime. To support this effort, automated tools are necessary to analyse changes and identify areas requiring high maintenance. One of these widely used tools in the industry is SonarQube. 
    
    In this paper, we highlighted some limitations of SonarQube and proposed a new tool, named SoHist, to address these limitations and perform historical code quality analysis. In addition, SoHist introduces a new visualization approach, which allows the weighting of code quality aspects according to individual project needs.
     
    The tool has demonstrated its benefits in two industrial settings, making it a good candidate for implementation in other industrial projects. Nevertheless, SoHist is still a prototype and thus has some limitations. At present, SoHist only shows the timestamp of the commits, but further details, such as the committer, still need to be included. Similarly, it is currently not possible to directly open the corresponding commit in GitLab or issue description in SonarQube. These usability issues may limit the overall root cause analysis of a project. Besides that, a potential risk to the long-term viability of SoHist is the possibility of restrictions imposed by SonarQube. 
     
    In future work, we will conduct further assessments to evaluate the usability, including performance and comprehensibility issues, in the industry domain. Furthermore, we plan to make improvements, such as automated plugin integration or disposal of current limitations. 
\begin{acks}
    This work has been supported by and done in the scope of the ITEA3 SmartDelta project, which has been funded by the Austrian Research Promotion Agency (Grant No. 890417) and  the German Federal Ministry of Education and Research (Grant No. 01IS21083A).
\end{acks}

\bibliographystyle{ACM-Reference-Format}
\bibliography{bibliography}


\begin{thebibliography}{13}


\ifx \showCODEN    \undefined \def \showCODEN     #1{\unskip}     \fi
\ifx \showDOI      \undefined \def \showDOI       #1{#1}\fi
\ifx \showISBNx    \undefined \def \showISBNx     #1{\unskip}     \fi
\ifx \showISBNxiii \undefined \def \showISBNxiii  #1{\unskip}     \fi
\ifx \showISSN     \undefined \def \showISSN      #1{\unskip}     \fi
\ifx \showLCCN     \undefined \def \showLCCN      #1{\unskip}     \fi
\ifx \shownote     \undefined \def \shownote      #1{#1}          \fi
\ifx \showarticletitle \undefined \def \showarticletitle #1{#1}   \fi
\ifx \showURL      \undefined \def \showURL       {\relax}        \fi
\providecommand\bibfield[2]{#2}
\providecommand\bibinfo[2]{#2}
\providecommand\natexlab[1]{#1}
\providecommand\showeprint[2][]{arXiv:#2}

\bibitem[Baldassarre et~al\mbox{.}(2020)]%
        {baldassarreDiffusenessTechnicalDebt2020}
\bibfield{author}{\bibinfo{person}{Maria~Teresa Baldassarre},
  \bibinfo{person}{Valentina Lenarduzzi}, \bibinfo{person}{Simone Romano},
  {and} \bibinfo{person}{Nyyti Saarim{\"a}ki}.}
  \bibinfo{year}{2020}\natexlab{}.
\newblock \showarticletitle{On the Diffuseness of Technical Debt Items and
  Accuracy of Remediation Time When Using {{SonarQube}}}.
\newblock \bibinfo{journal}{\emph{Information and Software Technology}}
  \bibinfo{volume}{128} (\bibinfo{date}{Dec.} \bibinfo{year}{2020}),
  \bibinfo{pages}{106377}.
\newblock
\showISSN{09505849}
\urldef\tempurl%
\url{https://doi.org/10.1016/j.infsof.2020.106377}
\showDOI{\tempurl}


\bibitem[Baskerville et~al\mbox{.}(2001)]%
        {baskervilleHowInternetSoftware2001}
\bibfield{author}{\bibinfo{person}{R. Baskerville}, \bibinfo{person}{L.
  Levine}, \bibinfo{person}{J. {Pries-Heje}}, {and} \bibinfo{person}{S.
  Slaughter}.} \bibinfo{year}{2001}\natexlab{}.
\newblock \showarticletitle{How {{Internet}} Software Companies Negotiate
  Quality}.
\newblock \bibinfo{journal}{\emph{Computer}} \bibinfo{volume}{34},
  \bibinfo{number}{5} (\bibinfo{date}{May} \bibinfo{year}{2001}),
  \bibinfo{pages}{51--57}.
\newblock
\showISSN{0018-9162}
\urldef\tempurl%
\url{https://doi.org/10.1109/2.920612}
\showDOI{\tempurl}


\bibitem[Buchwald(2022)]%
        {buchwaldHowOftenAre2022}
\bibfield{author}{\bibinfo{person}{Hendrik Buchwald}.}
  \bibinfo{year}{2022}\natexlab{}.
\newblock \bibinfo{title}{How Often Are Builtin Rules Updated in {{SQ}}}.
\newblock
\newblock
\urldef\tempurl%
\url{https://community.sonarsource.com/t/how-often-are-builtin-rules-updated-in-sq/69493}
\showURL{%
\tempurl}


\bibitem[Corrêa et~al\mbox{.}(2010)]%
        {5578300}
\bibfield{author}{\bibinfo{person}{Ulisses~Brisolara Corrêa},
  \bibinfo{person}{Luis Lamb}, \bibinfo{person}{Luigi Carro},
  \bibinfo{person}{Lisane Brisolara}, {and} \bibinfo{person}{Júlio Mattos}.}
  \bibinfo{year}{2010}\natexlab{}.
\newblock \showarticletitle{Towards Estimating Physical Properties of Embedded
  Systems using Software Quality Metrics}. In \bibinfo{booktitle}{\emph{2010
  10th IEEE International Conference on Computer and Information Technology}}.
  \bibinfo{publisher}{{IEEE}}, \bibinfo{address}{{Marrakech}},
  \bibinfo{pages}{2381--2386}.
\newblock
\urldef\tempurl%
\url{https://doi.org/10.1109/CIT.2010.409}
\showDOI{\tempurl}


\bibitem[Dornauer(2023a)]%
        {dornauerComputationsWeightedCode2023}
\bibfield{author}{\bibinfo{person}{Benedikt Dornauer}.}
  \bibinfo{year}{2023}\natexlab{a}.
\newblock \bibinfo{title}{Computations behind the {{Weighted Code Evolution
  Significance}}}.
\newblock
\newblock
\urldef\tempurl%
\url{https://doi.org/10.5281/ZENODO.7713698}
\showDOI{\tempurl}


\bibitem[Dornauer(2023b)]%
        {dornauerTrendCodeAnalysis2023}
\bibfield{author}{\bibinfo{person}{Benedikt Dornauer}.}
  \bibinfo{year}{2023}\natexlab{b}.
\newblock \bibinfo{title}{Trend of {{Code Analysis Tools}} 2004 -2023}.
\newblock
\newblock
\urldef\tempurl%
\url{https://doi.org/10.5281/ZENODO.7713953}
\showDOI{\tempurl}


\bibitem[Lenarduzzi et~al\mbox{.}(2020)]%
        {lenarduzziSurveyCodeAnalysis2020}
\bibfield{author}{\bibinfo{person}{Valentina Lenarduzzi},
  \bibinfo{person}{Alberto Sillitti}, {and} \bibinfo{person}{Davide Taibi}.}
  \bibinfo{year}{2020}\natexlab{}.
\newblock \showarticletitle{A {{Survey}} on {{Code Analysis Tools}} for
  {{Software Maintenance Prediction}}}.
\newblock In \bibinfo{booktitle}{\emph{Proceedings of 6th {{International
  Conference}} in {{Software Engineering}} for {{Defence Applications}}}},
  \bibfield{editor}{\bibinfo{person}{Paolo Ciancarini}, \bibinfo{person}{Manuel
  Mazzara}, \bibinfo{person}{Angelo Messina}, \bibinfo{person}{Alberto
  Sillitti}, {and} \bibinfo{person}{Giancarlo Succi}} (Eds.).
  Vol.~\bibinfo{volume}{925}. \bibinfo{publisher}{{Springer International
  Publishing}}, \bibinfo{address}{{Cham}}, \bibinfo{pages}{165--175}.
\newblock
\showISBNx{978-3-030-14686-3 978-3-030-14687-0}
\urldef\tempurl%
\url{https://doi.org/10.1007/978-3-030-14687-0_15}
\showDOI{\tempurl}


\bibitem[Rama{\v c} et~al\mbox{.}(2022)]%
        {ramacPrevalenceCommonCauses2022}
\bibfield{author}{\bibinfo{person}{Robert Rama{\v c}},
  \bibinfo{person}{Vladimir Mandi{\'c}}, \bibinfo{person}{Neboj{\v s}a Tau{\v
  s}an}, \bibinfo{person}{Nicolli Rios}, \bibinfo{person}{S{\'a}vio Freire},
  \bibinfo{person}{Boris P{\'e}rez}, \bibinfo{person}{Camilo Castellanos},
  \bibinfo{person}{Dar{\'i}o Correal}, \bibinfo{person}{Alexia Pacheco},
  \bibinfo{person}{Gustavo Lopez}, \bibinfo{person}{Clemente Izurieta},
  \bibinfo{person}{Carolyn Seaman}, {and} \bibinfo{person}{Rodrigo Spinola}.}
  \bibinfo{year}{2022}\natexlab{}.
\newblock \showarticletitle{Prevalence, Common Causes and Effects of Technical
  Debt: {{Results}} from a Family of Surveys with the {{IT}} Industry}.
\newblock \bibinfo{journal}{\emph{Journal of Systems and Software}}
  \bibinfo{volume}{184} (\bibinfo{date}{Feb.} \bibinfo{year}{2022}),
  \bibinfo{pages}{111114}.
\newblock
\showISSN{01641212}
\urldef\tempurl%
\url{https://doi.org/10.1016/j.jss.2021.111114}
\showDOI{\tempurl}


\bibitem[SonarSource(2023a)]%
        {sonarsourceMetricDefinition}
\bibfield{author}{\bibinfo{person}{SonarSource}.}
  \bibinfo{year}{2023}\natexlab{a}.
\newblock \bibinfo{title}{Metric Definition}.
\newblock
\newblock
\urldef\tempurl%
\url{https://docs.sonarqube.org/latest/user-guide/metric-definitions/}
\showURL{%
\tempurl}


\bibitem[SonarSource(2023b)]%
        {sonarsourceSonarQube}
\bibfield{author}{\bibinfo{person}{SonarSource}.}
  \bibinfo{year}{2023}\natexlab{b}.
\newblock \bibinfo{title}{{{SonarQube}}}.
\newblock
\newblock
\urldef\tempurl%
\url{https://www.sonarsource.com/products/sonarqube/}
\showURL{%
\tempurl}


\bibitem[SonarSource(2023c)]%
        {sonarsourceSonarQubeDownloads2023a}
\bibfield{author}{\bibinfo{person}{SonarSource}.}
  \bibinfo{year}{2023}\natexlab{c}.
\newblock \bibinfo{title}{{{SonarQube}} - {{Downloads}}}.
\newblock
\newblock
\urldef\tempurl%
\url{https://www.sonarsource.com/products/sonarqube/downloads/}
\showURL{%
\tempurl}


\bibitem[Spencer(2022)]%
        {spencerHaveSonarQubeCreate2022}
\bibfield{author}{\bibinfo{person}{Paul Spencer}.}
  \bibinfo{year}{2022}\natexlab{}.
\newblock \bibinfo{title}{Have {{SonarQube Create Historical Data}}}.
\newblock
\newblock
\urldef\tempurl%
\url{https://community.sonarsource.com/t/have-sonarqube-create-historical-data/60492}
\showURL{%
\tempurl}


\bibitem[Verner et~al\mbox{.}(2008)]%
        {vernerWhatFactorsLead2008b}
\bibfield{author}{\bibinfo{person}{June Verner}, \bibinfo{person}{Jennifer
  Sampson}, {and} \bibinfo{person}{Narciso Cerpa}.}
  \bibinfo{year}{2008}\natexlab{}.
\newblock \showarticletitle{What Factors Lead to Software Project Failure?}. In
  \bibinfo{booktitle}{\emph{2008 {{Second International Conference}} on
  {{Research Challenges}} in {{Information Science}}}}.
  \bibinfo{publisher}{{IEEE}}, \bibinfo{address}{{Marrakech}},
  \bibinfo{pages}{71--80}.
\newblock
\showISBNx{978-1-4244-1677-6}
\urldef\tempurl%
\url{https://doi.org/10.1109/RCIS.2008.4632095}
\showDOI{\tempurl}


\end{thebibliography}
\end{document}